\documentclass[preprint,showpacs,preprintnumbers]{revtex4}
\usepackage{amssymb}

\usepackage{amsmath}
\usepackage{graphicx}
\usepackage{dcolumn}
\usepackage{bm}

\begin{document}

\title{Absence of collapse in quantum Rabi oscillations}
\author{Shu He$^{1,2}$, Yang Zhao $^{3}$, and Qing-Hu Chen$^{1,2,*}$}

\address{
$^{1}$ Department of Physics, Zhejiang University, Hangzhou 310027,
P. R. China \\
$^{2}$ Center for Statistical and Theoretical Condensed Matter
Physics, Zhejiang Normal University, Jinhua 321004, P. R. China\\
$^{3}$  Division of Materials Science, Nanyang Technological
University, 50 Nanyang Avenue, Singapore 639798
 }
\date{\today }

\begin{abstract}
We show analytically that the collapse and revival in the population
dynamics of the atom-cavity coupled system under the rotating wave
approximation (RWA), valid only at very weak coupling, is an artifact as
the atom-cavity coupling is increased. Even the first-order
correction to the RWA is able to bring about the absence of the
collapse in the dynamics of atomic population inversion thanks to
intrinsic oscillations resulting from the transitions between two
levels with the same atomic quantum number. The removal of the
collapse is valid for a wide range of coupling strengths which are
accessible to current experimental setups. In addition, based on our
analytical results that greatly improve upon the conventional RWA,
even the strong-coupling power spectrum can now be explained with
the help of the numerically exact energy levels.
\end{abstract}


 \maketitle

\section{introduction}

The collapse and revival in the population dynamics of the
atom-cavity coupled system under the rotating wave approximation
(RWA) is an important issue in quantum optics \cite{Scully}. The
collapse is attributed to the destructive interference among various
transitions between the atomic upper and lower levels, and the
revival is due to quantized nature of the photonic number states in
the cavity. The collapse and revival phenomenon was first predicted
by Eberly {\it et al.}~\cite{eberly,eberly1}, and was observed
experimentally by Rempe {\it et al.}~\cite{rempe}.

The RWA is a very good approximation if the atom-cavity interaction
energy is much smaller than the characteristic energy of the
atom-cavity coupled system, i.e., if the ratio of the coupling
constant and the field frequency, $g/\omega$, is much less than
unity, a parameter space often refereed to as the weak coupling
regime. If the coupling constant is comparable to the field
frequency, i.e., $g/\omega \approx 0.1$, also known as the
ultra-strong coupling regime, the counter-rotating terms (CRTs) have
to be considered. The effect of the CRTs on the dynamics of the
atomic population inversion has been studied in the weak coupling
regime \cite{Muller,zubairy} as well as the deep strong coupling
regime with $g/\omega >1$ \cite{Casanova}. Zhang {\it et
al.}~examined the full range of atom-cavity coupling finding that
the collapse and revival gradually disappear and reemerge
periodically as the coupling strength increases in the strong
coupling regime \cite{Zhangyy}. Numerically exact calculations
reveal that the collapse and revival gradually loses prominence as
the coupling is increased, and vanish in the ultra-strong coupling
regime. However, due to the lack of an analytical apparatus
analogous to RWA, an in-depth understanding remains elusive on the
dynamics of atomic population inversion
from the weak to the ultra-strong coupling regime (%
$g/\omega \leq 0.2$).

For superconducting qubits, a one-dimensional transmission-line
resonator, or an LC circuit
\cite{Wallraff,Chiorescu,Wang,Fink,Deppe} can play the role of the
cavity, also known as the circuit quantum electrodynamics (QED)
systems. Recently, an LC resonator inductively coupled to a
superconducting qubit~\cite{Niemczyk,exp,Mooij} has been realized
experimentally. The
qubit-resonator coupling $g/\omega $ has been strengthened to $\thicksim 0.1$%
, entering the ultra-strong coupling regime, and evidence for the
breakdown of the RWA has been provided~\cite{Niemczyk}. Therefore,
the CRTs can no longer be ignored. Recently, much attention has been
devoted to the qubit-oscillator system and the effect of the CRTs in
the ultra-strong coupling regime \cite
{Werlang,Hanggi,Nori,Hausinger,chen10b,Zhang13}.

A two-level atom coupled to a single-mode cavity has long been a
subject of extensive study, for which two main schemes are usually
employed. One is based on the phontonic Fock states as pioneered by
Swain \cite {Swain,Kus,Durstt,Tur,Bishop}, and the other, on various
unitary  transformations or displaced oscillators \cite
{Feranchuk,Irish,chenqh,liu,chen10,zheng,QingHu1,Braak,exp,Beaudoin,Chen2012a,zhao94,wu2013},
which are equivalent to extended coherent states
\cite{chen94} or extended squeezed states~\cite{Nori}. Very accurate solutions can be obtained in the latter
scheme, but as an infinite number of phontonic Fock states are
involved, certain important features may be smeared out.

Recently, He {\it et al.}~proposed a method of using only the
dominant parts in Swain's wave function \cite{Chen2012b} in the
corrections to the RWA. The effect of the CRTs emerges clearly even
in the first-order correction. All eigenvalues and eigenfunctions
are derived analytically. The vacuum Rabi splitting has been
obtained successfully up to the remarkable coupling strength of
$g=0.4$, suggesting that they could be convincingly applied to
recent circuit QED systems operating in the ultra-strong coupling
regime. In this work, we show that this method can be improved
further with easy-to-use eigenvalues and eigenvectors similar to
those in the RWA, and the atomic population inversion from an
initial field of coherent states can then be studied in great
detail.

The paper is organized as follows. In Section II, we show an
improved version of the first correction to the RWA, and give all
eigensolutions similar as those in RWA. In Section III, the atomic
population inversion is calculated analytically. The origin of the
absence of the collapse up to the ultra-strong coupling regime is
examined in detail. The structure of the exact power spectrum is
discussed in terms of the corrections to the RWA and the numerical
exact energy levels. A brief summary is given in Section IV.

\section{Corrected rotating-wave approximations}

The Hamiltonian of a two-level atom (a qubit) with transition frequency $%
\Delta $ interacting with a single-mode quantized cavity of frequency $%
\omega $ takes the form
\begin{equation}
H=\frac{\Delta }{2}\sigma _{z}+\omega a^{\dagger }a+g\left( a^{\dagger
}+a\right) \sigma _{x},
\end{equation}
where $g$ is coupling strength, $\sigma _{x}$ and $\sigma _{z}$ are
Pauli spin-$1/2$ operators, and $a^{\dagger }$ and $a$ are the
creation and annihilation operators for the quantized field. Here we
define $\delta =\Delta -\omega $ as the  detuning
parameter, and focus on the resonance case of $\delta =0$. The
frequency of the cavity mode is set to unity as the energy scale,
i.e., $\omega =1$.

The RWA neglects the CRTs, $a^{\dagger }\sigma _{+}+a\sigma _{-}$, rendering
the Hamiltonian diagonalizable with the eigenfunctions~\cite{Scully}
\begin{equation}
\left| kn\right\rangle ^{\mathrm{RWA}}=\frac 1{\sqrt{2}}\left( \
\begin{array}{l}
\left( -1\right) ^k\;\left| n\right\rangle \\
\ \left| n+1\right\rangle
\end{array}
\right) ,~k=1,2,~n=0,1,2,...  \label{waveRWA}
\end{equation}
and the corresponding eigenvalues
\begin{equation}
E_{kn}^{\mathrm{RWA}}=n+\frac 12+\left( -1\right) ^k\sqrt{n+1}g,
\label{EknRWA}
\end{equation}
where $k=1,2$\ is the atomic quantum number, and $n=0,1,2,...$ is the photonic
quantum number. Throughout this appear, $n$ and $k$ are regarded as the
level indices.

A unified expression in the first-order correction to the RWA was
proposed for the eigenvalues and eigenfucntions in a recent work
\cite{Chen2012b}, which recovers the RWA expression in the absence
of the correction. In this paper, we employ an improved selection
rule for the roots of the derived univariate cubic equation, and
obtain a modified general expression with details given below.

The first correction to the RWA wave function is to add a new
photonic state $\left\vert n+2\right\rangle $ to the upper atomic
level so that the wave function can be written as
\begin{equation}
\left\vert kn\right\rangle =\left( \
\begin{array}{l}
c_{kn}^{(0)}\left\vert n\right\rangle +c_{kn}^{(2)}\left\vert
n+2\right\rangle \\
\;\;\;\;\;c_{kn}^{(1)}\left\vert n+1\right\rangle
\end{array}
\right) .  \label{waveCRWA}
\end{equation}
From the RWA results, we note that an odd quantum number $n$ is for
even parity, and an even $n$ for odd parity, resulting in a single
univariate cubic equation
\[
E^{3}-\left( 3n+\frac{7}{2}\right) E^{2}+\left[ \left( n+\frac{1}{2}\right)
\left( 3n+\frac{11}{2}\right) -\left( 2n+3\right) g^{2}\right] E
\]
\begin{equation}
-\left( n+\frac{1}{2}\right) ^{2}\left( n+\frac{5}{2}\right) +\left(
2n^{2}+6n+\frac{7}{2}\right) g^{2}=0,  \label{base}
\end{equation}
which is the same as Eq.~(22) in Ref.~\cite{Chen2012b}. In Sec.~III
(B)1 of Ref.~\cite{Chen2012b}, however, a scenario of an even
quantum number $n$ with even parity and an odd $n$ with odd parity
was also considered, which does not arise in Hamiltonian (1) where
only one univariate cubic equation is needed to account for all
physical results. Note that Eq.~(\ref{base}) in principle gives
three roots. Comparing with the RWA results, we know that each $n$
corresponds to two eigenvalues, one of which is superfluous and
should be omitted. By fitting the numerically exact results, two
roots are chosen:
\begin{equation}
E_{kn}=\frac{\left( 3n+\frac{7}{2}\right) +\sqrt{\allowbreak \left(
6n+9\right) g^{2}+4}\left[ \cos \theta +\left( -1\right) ^{k}\sqrt{3}\sin
\theta \right] }{3},  \label{EknCRWA}
\end{equation}
where
\[
\theta =\frac{1}{3}\arccos \left( \frac{-8+9ng^{2}}{\sqrt{\left[ 4+\left(
6n+9\right) g^{2}\right] ^{3}}}\right) +\frac{2\pi }{3},
\]
$\allowbreak $ and the ratio of coefficients in the eigenfunctions (\ref
{waveCRWA}) is
\begin{equation}
c_{kn}^{(0)}:c_{kn}^{(1)}:c_{kn}^{(2)}=\left[ -\frac{\sqrt{n+1}}{\Omega
_{n}(E_{kn})}\right] :1:\left[ -\frac{\sqrt{n+2}}{\Omega _{n+2}(E_{kn})}%
\right] ,  \label{coeff_CRWA}
\end{equation}
with
\[
\Omega _{n}(E_{kn})=\frac{1}{g}\left[ n-E_{kn}+\frac{\Delta }{2}\right] .
\]
Eqs.~(\ref{waveCRWA}) and (\ref{EknCRWA}) are the counterparts of RWA
results, Eqs.~(\ref{waveRWA}) and (\ref{EknRWA}).

It is easily seen that the leading wave function correction
$\left\vert \uparrow, n+2\right\rangle $ in the upper atomic level is
produced by the CRTs, $a^{\dagger }\sigma _{+}+a\sigma _{-}$. This
first correction to RWA, which will be denoted as CRWA, is
responsible for numerous physical processes beyond the RWA, and with
eigenvalues and eigenvectors explicitly given, corresponding to the
RWA results one by one, applications of CRWA can be conveniently
made.

To better compare with the RWA, the CRWA eigenvalues can be expanded in
terms of the coupling constant $g$ as follows
\begin{equation}
E_{kn}=n+\frac 12+\left( -1\right) ^k\sqrt{n+1}g-\frac{n+2}4g^2-\left(
-1\right) ^k\frac{\left( n+2\right) \left( 3n+2\right) }{32\sqrt{n+1}}%
g^3+O(g^3),  \label{energy_expa}
\end{equation}
and the three coefficients in the normalized eigenfunctions read
\begin{align}
& c_{kn}^{(0)}=(-1)^k\frac{\sqrt{2}}2+\frac{\sqrt{2}(n+2)}{16\sqrt{n+1}}%
g-(-1)^k\frac{\sqrt{2}(n+2)^2}{256(n+1)}g^2+O(g^2),  \nonumber \\
& c_{kn}^{(1)}=\frac{\sqrt{2}}2-(-1)^k\frac{\sqrt{2}(n+2)}{16\sqrt{n+1}}g-%
\frac{\sqrt{2}(n+2)(17n+18)}{256(n+1)}g^2+O(g^2),  \nonumber \\
& c_{kn}^{(2)}=-\frac{\sqrt{2n+4}}4g-(-1)^k\frac{(3n+2)\sqrt{2n+4}}{32\sqrt{%
n+1}}g^2+O(g^2).  \label{coeff_expa}
\end{align}
Comparing Eq.~(\ref{energy_expa}) with the RWA counterpart Eq.~(\ref{EknRWA}%
), it is found that the conventional RWA results are recovered if
only 1st-order (in $g$) terms are kept in Eq.~(\ref{energy_expa}). A
similar comparison between Eq.~(\ref{coeff_expa}) and
Eq.~(\ref{waveRWA}) reveals that the RWA eigenstates appear
explicitly in the expressions of the CRWA
counterparts. It makes sense that the coefficients $c_{kn}^{(0)}$and $%
c_{kn}^{(1)}$ of $\left| \uparrow n\right\rangle $ and$\ \left| \downarrow
n+1\right\rangle \;$in the RWA states, respectively, are of order $1$, and
coefficients$\ c_{kn}^{(2)}$ for the new state $\left| \uparrow
n+2\right\rangle $ are of order $g$. It will be demonstrated that the CRWA
expansions are exact up to $g^2$ ($g$) in energy (wave function). Here we
keep one higher order term in both the eigenvalues and eigenvectors, which
can be shown to reproduce Eqs.~(\ref{EknCRWA}) and (\ref{waveCRWA}) with
sufficient accuracy, and will therefore be used to calculate all CRWA
results in this work. As the CRWA eigenvectors are approximate, they are not
all strictly orthogonal. For example, for $k_1\neq k_2,$ $\left\langle
k_1n\right| \left| k_2n\right\rangle \propto g^4\neq 0$. It is understood
that orthogonality is only observed in the weak coupling limit.

Following the standard procedure, we can also obtain the ground state (GS)
and its energy by adding one new bare state to the RWA one $\left\vert
\downarrow 0\right\rangle $ as
\begin{equation}
\left\vert GS\right\rangle =\left( \
\begin{array}{l}
\;d_{1}\left\vert 1\right\rangle \; \\
\;d_{0}\left\vert 0\right\rangle
\end{array}
\right),
\end{equation}
where
\begin{eqnarray}
d_{0} &=&\allowbreak 1-\frac{1}{8}g^{2}+\frac{11}{128}g^{4},  \nonumber \\
d_{1} &=&-\frac{1}{2}g+\frac{3}{16}g^{3},  \nonumber
\end{eqnarray}
and
\begin{equation}
E_{GS}=-\frac{1}{2}-\frac{1}{2}g^{2}+\frac{1}{8}g^{4}.
\end{equation}
This GS eigenvector is regarded as the CRWA one, because only one new
phontonic state is added, analogous to the CRWA excited states.

Similarly, we can also obtain the GS energy and state with higher order
corrections
\begin{eqnarray}
E_{GS}^{(2)} &=&-\frac 12-\frac 12g^2-\frac 18g^4+O\left( g^6\right) ,
\label{SGS} \\
\left| GS\right\rangle ^{(2)} &=&\left( \
\begin{array}{l}
\;\;\;\;d_1^{(2)}\left| 1\right\rangle \\
d_0^{(2)}\left| 0\right\rangle +d_2^{(2)}\left| 2\right\rangle
\end{array}
\right) ,
\end{eqnarray}
with
\begin{eqnarray*}
d_0^{(2)} &=&1-\frac 18g^2-\frac{13}{128}g^4+O\left( g^6\right) , \\
d_1^{(2)} &=&-\allowbreak \frac 12g-\frac 1{16}g^3+O\left( g^5\right) , \\
d_2^{(2)} &=&\frac{\sqrt{2}}4g^2-\frac{\sqrt{2}}{32}g^4+O\left( g^6\right) .
\end{eqnarray*}
It is interesting to note that the eigenvalues (eigenvectors) up to $g^2$ ($%
g $) in the CRWA are not modified in the second corrections of the GS state.
The GS eigenvectors appear to reveal the general property of the excited
states with second order corrections, for which analytical expressions are
elusive.

CRWA may not yield more accurate results than many other approaches  involving infinitely
many photon states. The advantage, instead, lies in
 the transparency in important mechanisms of interest provided by
the compact eigenstates with only three photonic number states.
Now with all CRWA eigenvalues and eigenvectors at hand, we can revisit many
physical problems that have been studied using the RWA. In this paper, we
focus on the quantum dynamical effects, one fundamental issue in quantum
optics.

\section{atomic population inversion from coherent state}

\subsection{The CRWA Results}

We first consider a two-level atom interacting with a coherent field. The
initial coherent state in the upper level can be expanded in terms of the
photonic number states
\begin{equation}
\left| \alpha \right\rangle =\sum_n\beta _n\left| \uparrow ,n\right\rangle ,
\label{coherent}
\end{equation}
where $\beta _n=\exp \left( -\left| \alpha \right| ^2/2\right) \frac{\alpha
^n}{\sqrt{n!}}$ is a Poisson distribution , $\alpha ^2=\overline{n}\;$ is
the average photon number.

\begin{figure}[tbp]
\includegraphics[width=12cm]{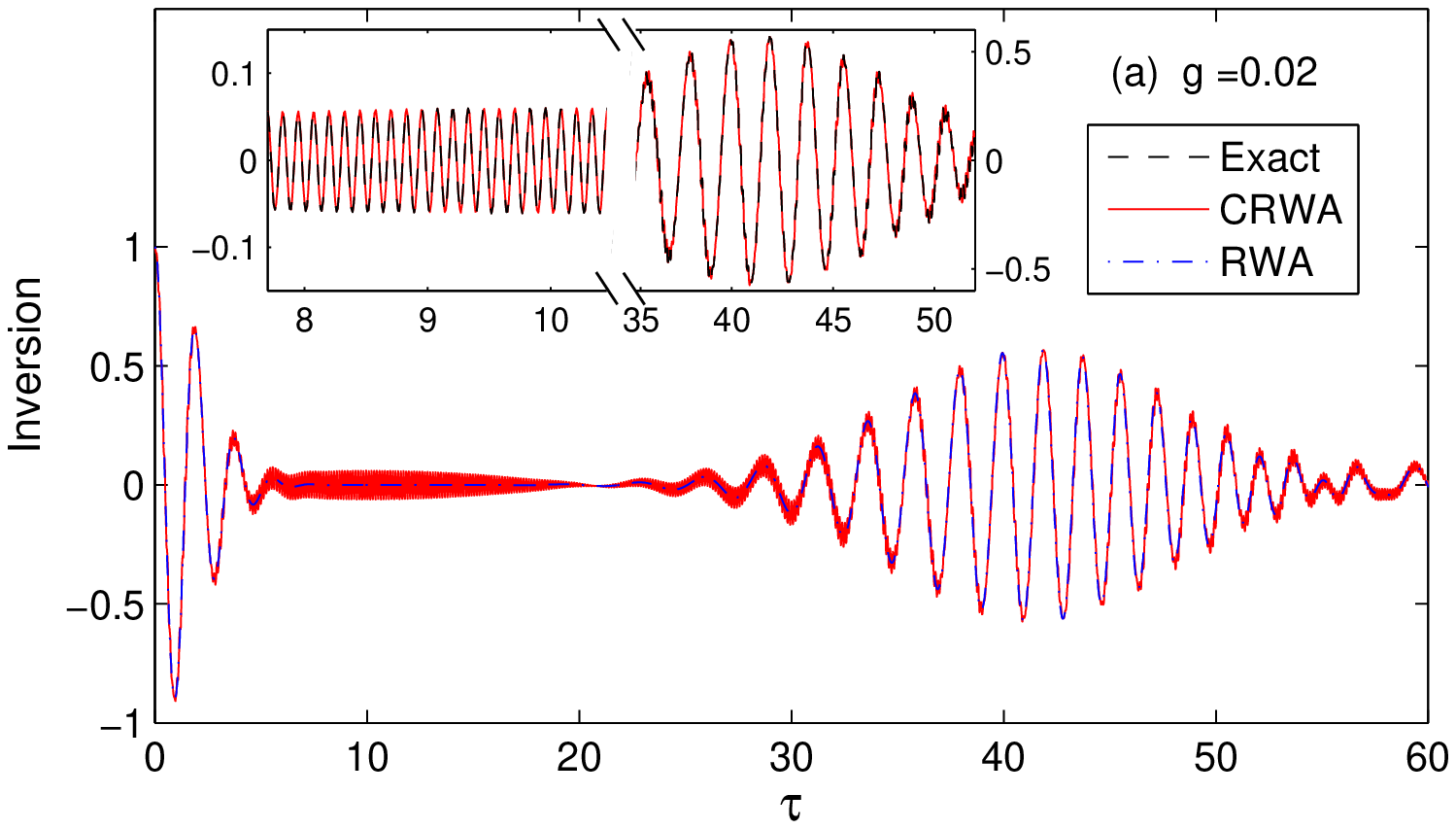} %
\includegraphics[width=12cm]{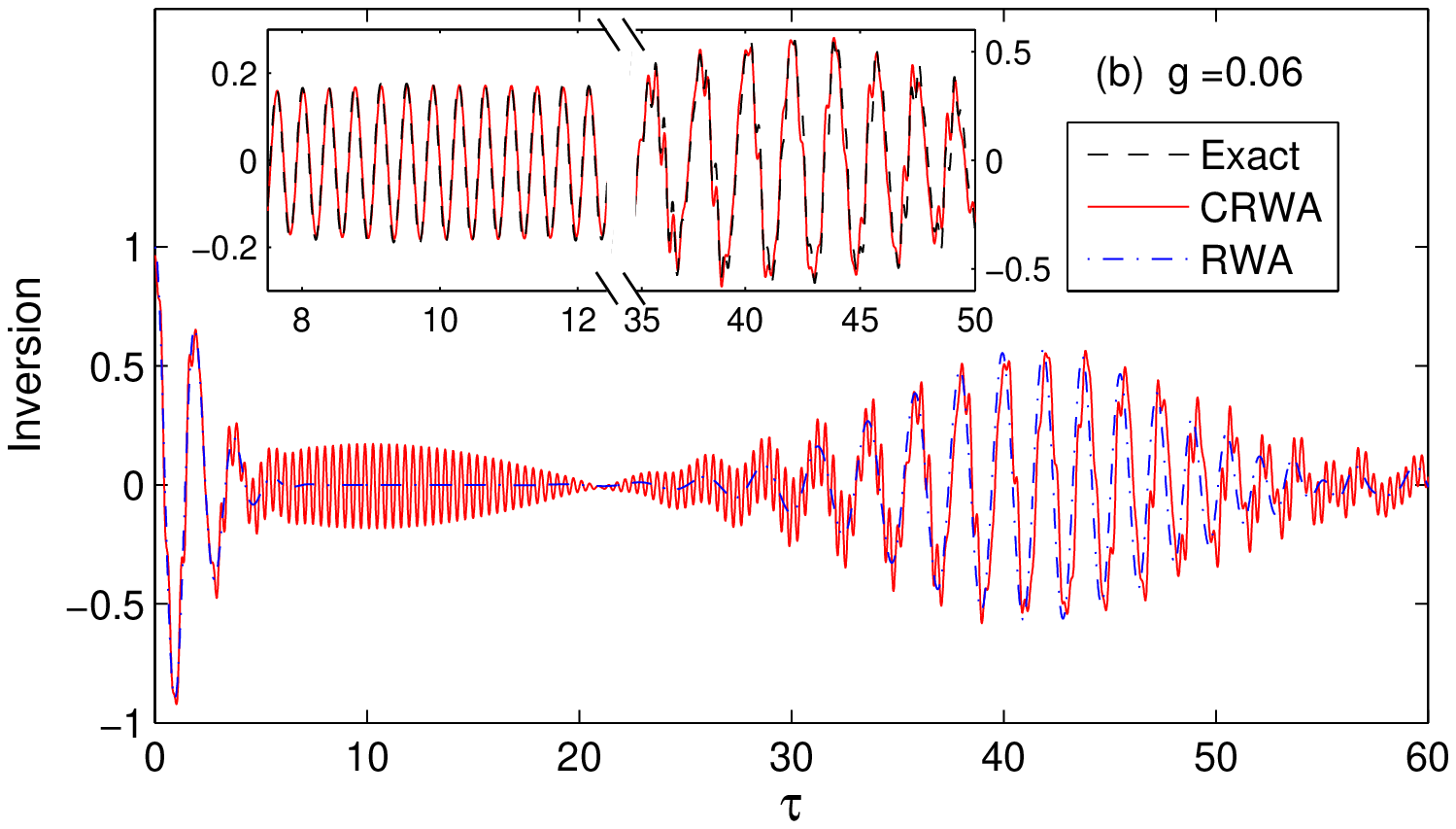} %
\caption{ (Color online) The atomic population inversion for an
initial coherent state for $g=0.02$ and $0.06$ using  the CRWA
(solid red), the RWA ( dashed-dotted blue), and numerically exact
calculations ( dashed black) with an average photon number $\alpha
^{2}=10$. The reduced time $\tau=2gt$.} \label{collapse}
\end{figure}

It is well known that the atomic population inversion exhibits
collapse followed by periodic revivals in the RWA~\cite{Scully}. It
is intriguing to
ask what would happen beyond the RWA. Using the CRWA results given in Eqs.~(%
\ref{energy_expa}) and (\ref{coeff_expa}), one can easily calculate
the dynamics of the atomic population inversion. The results are presented in Fig.~%
\ref{collapse} for $\alpha ^{2}=10$ and two coupling strengths $%
g=0.02$ and $0.06$. The corresponding RWA results are also shown in
Fig.~\ref{collapse} with differences accounted for by the CRTs. To
demonstrate the validity of the CRWA, we compare in the inset of Fig.~\ref{collapse}
the numerically exact
results with the CRWA counterparts in two time periods of large oscillation amplitudes. Surprisingly, our CRWA results are in
excellent agreement with the exact ones up to the ultra-strong
coupling regime (with $g$ in the order of $0.1$). Actually, the
agreement can be kept up to $g=0.2$ (not shown here), a coupling
strength almost doubling the current experimentally accessible
value.

Interestingly, a new oscillation appears during the collapses with a
coupling-independent frequency and an amplitude roughly proportional
to the coupling strength. This is to say, the well-known collapse
observed in the RWA is absent due to
the presence of the CRTs, even at a very weak coupling strength (such as $%
g=0.02$), where the RWA is usually considered to be valid. While it
is difficult for certain accurate approaches based on an infinite
series of number states to uncover this mysterious phenomenon, here
we will derive the atomic population inversion analytically and
discuss the detailed dynamics based on the CRWA with only one new
state added to the RWA one.

\subsection{Detailed process in terms of CRWA}

We can expand the initial coherent state in the upper level in terms
of the CRWA eigenstates, and then obtain the atomic population
inversion. We list here individual contributions for the benefit of
discussion leaving detailed derivations to Appendix A:
\begin{equation}
W(t)=\left( 2C-1\right) +2\sum_{k=1,2}S_k\cos [(E_{GS}-E_{k1})t]+W^{(\mathrm{%
Rabi})}(t)+W_{sk}^{(I)}(t)+W_{dk}^{(I)}(t),  \label{AI_CRWA}
\end{equation}
where
\begin{eqnarray}
&&W^{(\mathrm{Rabi})}(t)=2\sum_{n=0}^\infty R_n\cos \left[
(E_{2n}-E_{1n})t\right] ,  \label{AI(1)} \\
W_{sk}^{(I)}(t) &=&2\sum_{n=0}^\infty \left\{ I_{1n}\cos \left[ \left(
E_{1n}-E_{1n+2}\right) t\right] +I_{2n}\cos \left[ \left(
E_{2n}-E_{2n+2}\right) \right] t\right\} ,  \label{AI(2)} \\
W_{dk}^{(I)}(t) &=&2\sum_{n=0}^\infty \left\{ I_{12n}\cos \left[
(E_{1n}-E_{2n+2})t\right] +I_{21n}\cos \left[ (E_{2n}-E_{1n+2})t\right]
\right\} ,  \label{AI(3)}
\end{eqnarray}
where all coefficients are given in Eqs.~(\ref{coeff_final}) in
Appendix A. In Eq.~(\ref{AI_CRWA}), the first constant term and the
second terms for transitions to the GS state are extremely small for
$g\leq 0.2$, and can be omitted. The main contributions to the
atomic population inversion are the
three kinds of oscillations given by Eqs. (\ref{AI(1)}), (\ref{AI(2)}), and (%
\ref{AI(3)}). Eq.~(\ref{AI(1)}) depicts the modified Rabi oscillations,
which are not essentially different from the usual one in the RWA. Eqs.~(\ref
{AI(2)}) and (\ref{AI(3)}) describe the transitions between the ($k,n$) and ($k,n+2$)
levels, and those between the ($k,n$) and ($%
k^{\prime }\neq k,n+2$) levels, respectively.

\begin{figure}[tbp]
\includegraphics[width=10cm]{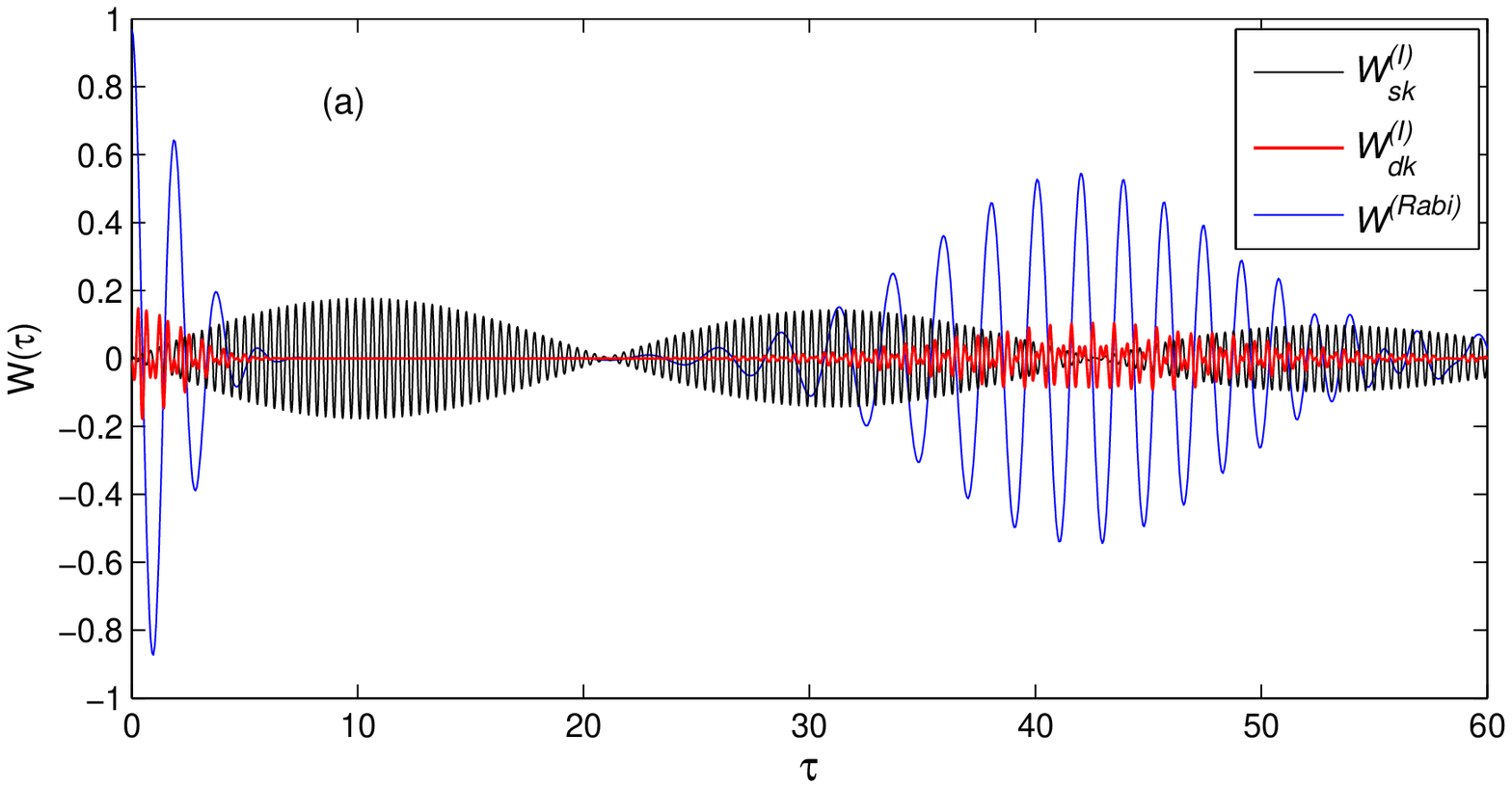} %
\includegraphics[width=10cm]{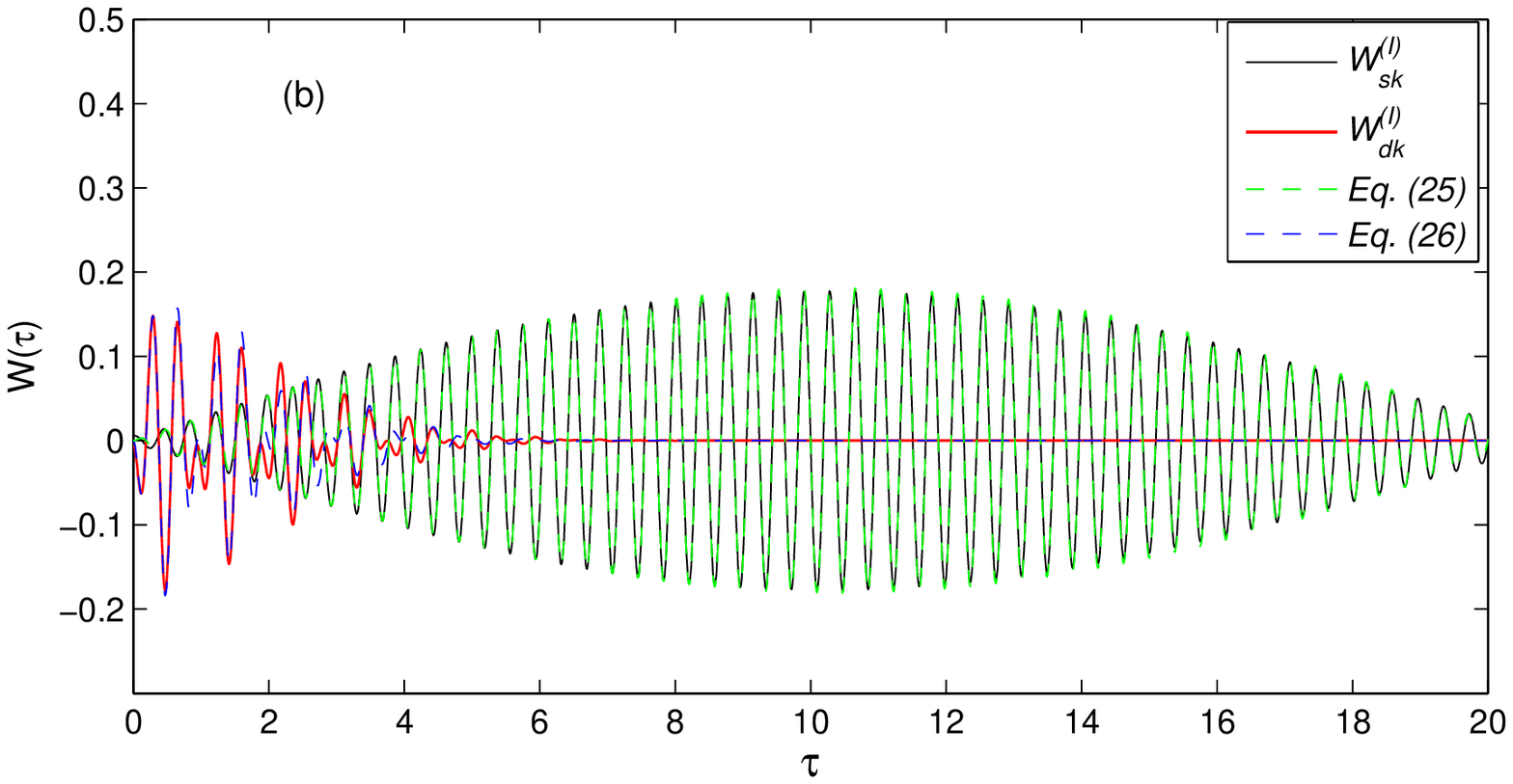}
\caption{ (Color online) (a) Various contributions to the time dependent
atomic population inversion: $W^{(I)}_{sk}$ for transitions from the levels in
different $n$ with the same $k $ , and $W^{(I)}_{dk}$ with different $k $,
modified Rabi oscillations $W^{(\mathrm{Rabi})}(t)$ at $g=0.06$, $\alpha
^{2}=10$. (b) Comparisons with the approximate short-time dynamics given in
Eqs.~(\ref{I1_a}) and (\ref{I2_a}). The reduced time $\tau=2gt$. }
\label{g006}
\end{figure}

To show contributions individually, we calculate Eqs. (\ref{AI(1)}), (\ref
{AI(2)}), and (\ref{AI(3)}) independently. Fig.~\ref{g006} (a) shows the
three contributions for $g=0.06$. The black lines depict the atomic
population inversion for the transitions between the ($k,n$) and ($k,n+2$) levels,
revealing the absence of the usual collapse. For the transitions between the ($%
k,n $) and ($k^{\prime }\neq k,n+2$) levels, the atomic population
inversion, as shown by the red lines, displays the collapse and
revivals as in the Rabi oscillations. The modified Rabi oscillations
are shown as the blue lines. The main results in Fig.~1(b) can then
be completely reproduced by summing up these three contributions.

To better illustrate the point, we collect the amplitude terms up to order $%
g $ and the energy difference terms up to order $g^2$ in Eq.~(\ref{w_tot}),
obtaining a concise expression for the atomic population inversion:
\begin{equation}
W(t)=\sum_{n=0}^\infty \beta _n^2\left[ \cos \left[ (E_{2n}-E_{1n})t\right] -%
\frac{g\alpha ^2}{2\sqrt{n+1}}\sum_{k,k^{\prime }=1,2}\left( -1\right)
^{k^{\prime }}\cos \left[ \left( E_{k,n+2}-E_{k^{\prime },n}\right) t\right]
\right] ,  \label{CRWA0}
\end{equation}
where
\begin{align}
& E_{2n}-E_n=2\sqrt{n+1}g,  \label{Rabi_E} \\
& E_{k,n+2}-E_{k,n}=2-\frac{g^2}2+\left( -1\right) ^kg\left( \sqrt{n+3}-%
\sqrt{n+1}\right) ,  \label{I_close} \\
& E_{k,n+2}-E_{k^{\prime }\neq k,n}=2-\frac{g^2}2+\left( -1\right)
^kg\left( \sqrt{n+3}+\sqrt{n+1}\right) .  \label{I_apart}
\end{align}
The first term in Eq.~(\ref{CRWA0}) is the conventional Rabi
oscillation, which can be approximated by $\cos \left(
2gt\sqrt{\alpha ^2+1}\right) e^{-\frac 12\left( gt\right) ^2}$ at
short times ($gt<\alpha $). The collapse occurs at
$t_{\mathrm{collapse}}\thicksim 1/g$ due to the exponential factor,
and the usual Rabi frequency is $2g\sqrt{\alpha ^2+1}$. The second
term in Eq.~(\ref{CRWA0}) includes four oscillations. All are from
the transitions between $(k, n)$ and $(k, n+2)$ levels, which, despite
being ubiquitous in exact dynamics even for weak coupling, are
absent in the RWA. In fact, the inter-level oscillations, also known
as the intrinsic oscillations, are even captured by semi-classical
descriptions of two-level atoms coupled with a classical field.

Now we focus on the energy difference, i.e., the frequency of the
oscillations. It is observed that the energy differences with the same $k$
[Eq.~(21)] are very small, and those with the different $k$ [Eq.~(22)] are
large. Regrouping the summation in the second term of Eq.~(\ref{CRWA0})
yields
\begin{equation}
W_{sk}^{(I)}(t)=g\alpha ^2\sin \left[ \left( 2-\frac 12g^2\right) t\right]
\sum_{n=0}^\infty \frac{\beta _n^2}{\sqrt{n+1}}\sin \left[ g\left( \sqrt{n+3}%
-\sqrt{n+1}\right) t\right] ,  \label{I1}
\end{equation}
\begin{equation}
W_{dk}^{(I)}(t)=-g\alpha ^2\sin \left[ \left( 2-\frac 12g^2\right) t\right]
\sum_{n=0}^\infty \frac{\beta _n^2}{\sqrt{n+1}}\sin \left[ g\left( \sqrt{n+1}%
+\sqrt{n+3}\right) t\right] .  \label{I2}
\end{equation}
Note that they share a common fast oscillation with a weakly $g$-dependent
frequency of $2-\frac 12g^2$. However, their envelopes are given by
different infinite summations.

Considering the weight $\beta _n$ in the summation, the dominate term in the
envelope of Eq.~(\ref{I1}) is from the $n=\alpha ^2$ term, so the short-time
dynamics is roughly described by
\begin{equation}
W_{sk}^{(I)}(t)\thickapprox \frac{g\alpha ^2}{\sqrt{\alpha ^2+1}}\sin \left[
\left( 2-\frac 12g^2\right) t\right] \sin \left[ \frac{gt}{\sqrt{\alpha ^2+1}%
}\right] .  \label{I1_a}
\end{equation}
As shown in Fig.~\ref{g006} (b) by the green line, Eq.~(\ref{I1}) describes
the short-time dynamics very well. It is well known that the first collapse
under the RWA persists up to the revival time $\tau _{\mathrm{revival}}=2\pi
\alpha $. For $\alpha ^2>>1$, the argument in the second sin function in
Eq.~(\ref{I1_a}) at the revival time is about $\pi $. So in the RWA collapse
regime, this fast oscillation in the CRWA is enveloped sinusoidally with a
half period with a maximum amplitude of about $g\alpha $. For the given
parameters, the first revival time is $\tau _{\mathrm{revival}}\thickapprox
20$, and the half period for the first envelope should be also approximately
$20$, which is just in excellent agreement with Fig.~\ref{g006} (b).
Actually for any average photonic number, there is no finite collapse regime
for the fast oscillations in the short-time dynamics. It follows that there
is absolutely no collapse in the absence of RWA.

While the other intrinsic oscillations described by Eq.~(\ref{I2}) at
short times can be approximated by
\begin{equation}
W_{dk}^{(I)}(t)\thickapprox -g\alpha \sin \left[ \left( 2-\frac 12g^2\right)
t\right] \sin (2g\alpha t)\exp \left[ -\frac{\left( gt\right) ^2}2\right].
\label{I2_a}
\end{equation}
The detailed derivation is presented in Appendix B. It is also demonstrated
by the blue line in Fig.~\ref{g006} (b) that the above approximation can
capture the short-time dynamics quite well. Note that the intrinsic
oscillations from the transitions between ($k,n$) and ($k^{\prime }\neq k,n+2
$) levels also collapse at the same time as the Rabi
oscillations thanks to the same decay factor $\exp \left[ -\left( gt\right)
^2/2\right] $ such that they do not contribute to the absence of the collapse
at all.

To sum up, the transitions between levels of $n$ and $n+2$ with the same $k $
bring about a fast intrinsic oscillation with amplitude $g\alpha $, which is
remarkable so that the collapse actually never occur in the real system.
Previous collapse is only an artificial results from the theoretical RWA. As
exhibited in Fig. 1  even for $g=0.02$, a visible intrinsic oscillation
appears in the usual RWA collapse regime. As the coupling strength in the
current circuit QED systems has reached $g=0.1$, the absence of collapse can
be checked experimentally.

We believe that a first order perturbation technique in the path
integral framework \cite{zubairy} would give similar results to
order $g$ at weak coupling. Their findings were attributed to the
CRTs as a whole, and mechanisms relevant to the transitions are
elusive due to the fact that detailed information on the eigenstates is
not available in the path integral approach which integrates out all
bosonic degrees of freedom. In addition, compared with the exact
solution, our CRWA results may not be better than the second-order perturbative one, but an
infinite number of phontonic Fock states are involved in the second-order perturbation theory,
depriving its analytical clarity.

\subsection{Power spectrum}

To probe further various oscillations in the atomic population inversion, we
calculate the power spectrum $F(\omega )$ defined as follows
\begin{equation}
F(\omega )=\left| \int_0^\infty W(t)\exp \left( -i\omega \ t\right)
dt\right| ^2.
\end{equation}
The power spectrum will be calculated exactly and analyzed using our CRWA
results, in which various transitions can be identified separately.

\begin{figure}[tbp]
\includegraphics[width=10cm]{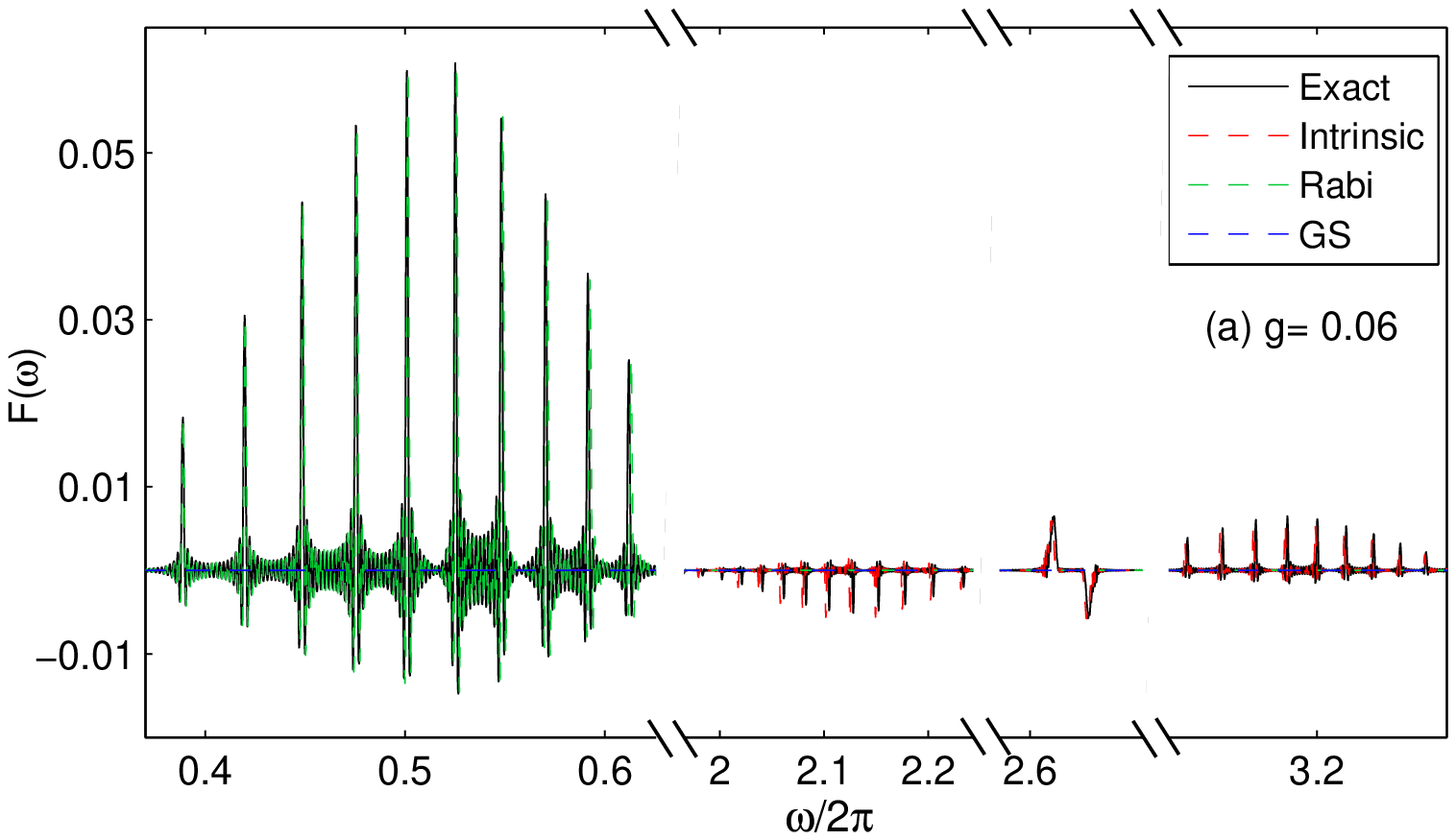} %
\includegraphics[width=10cm]{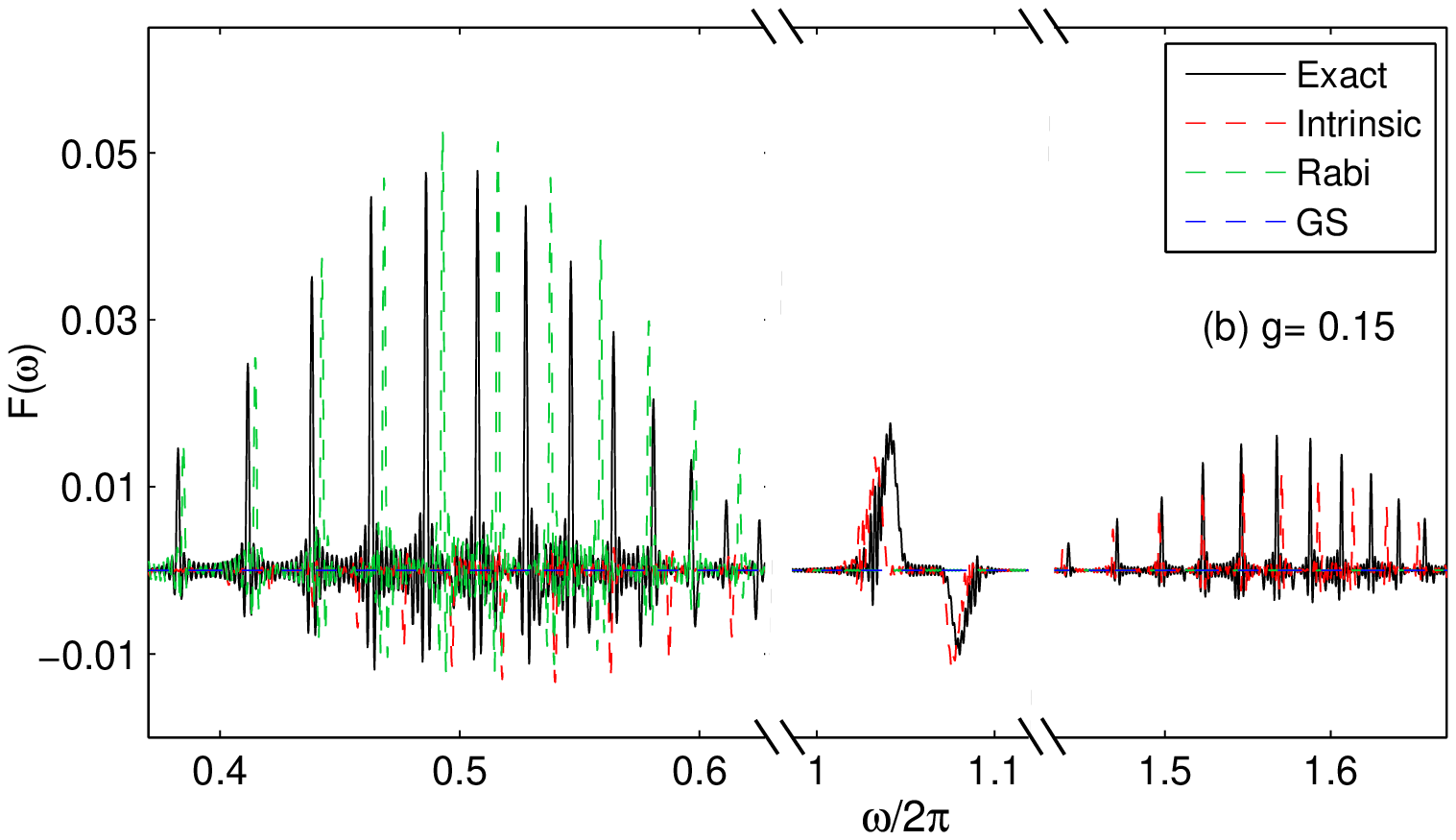} %
\includegraphics[width=10cm]{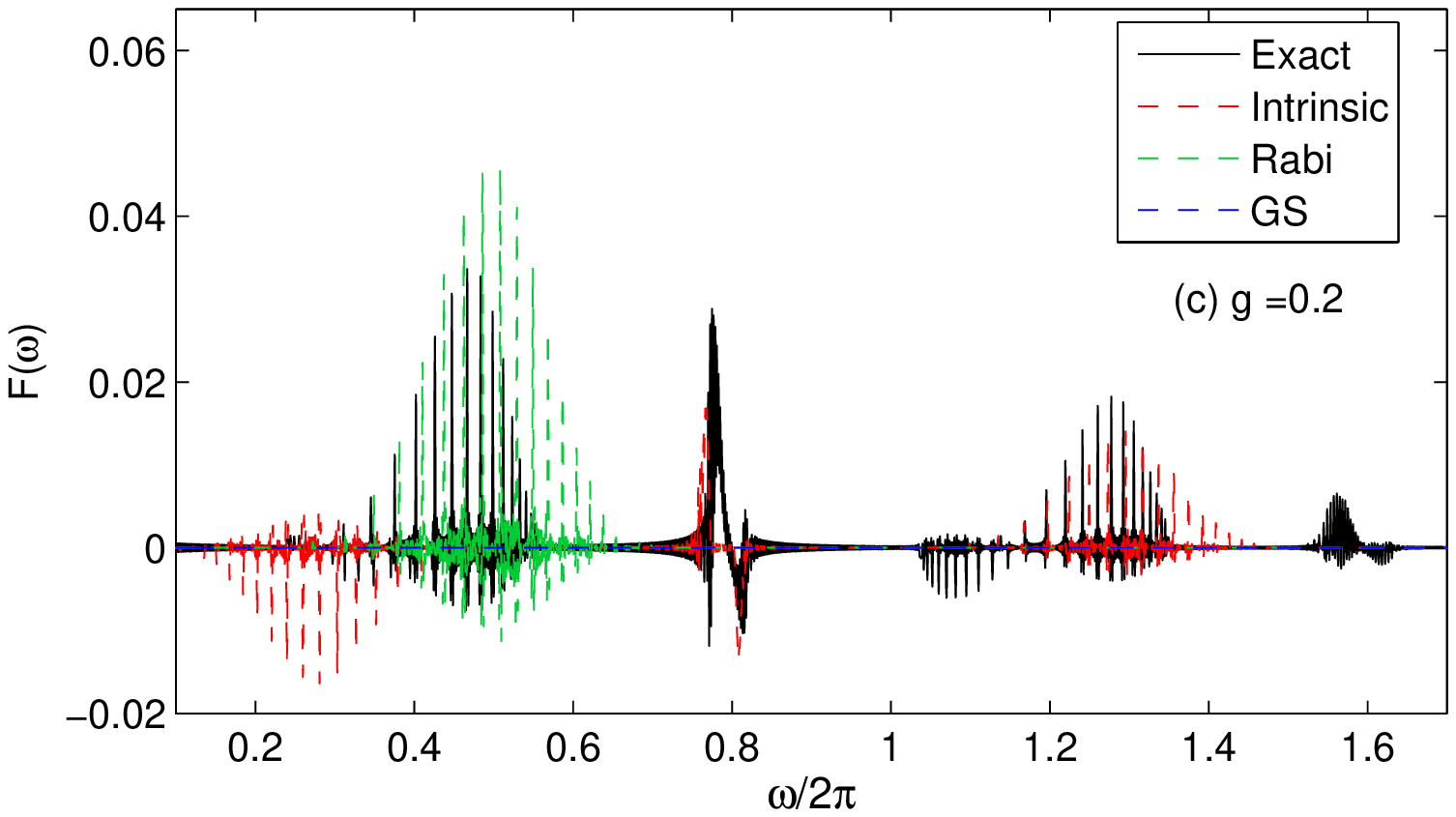}
\caption{ (Color online) The power spectrum for $g=0.06, 0.15$, and $0.2$
with CRWA and numerical exact study at average photon number $\alpha ^{2}=10$%
. The frequency unit is $2g$.}
\label{spectrum}
\end{figure}

In Fig.~\ref{spectrum}, the numerically exact power
spectrum for various coupling strengths are given by
solid black lines for the case of $\alpha ^2=10$. We also list the power spectrum individually
from different time-dependent components such as the Rabi
oscillations (green lines) and intrinsic oscillations (red lines) in
Eq.~(\ref{CRWA0}). It is very interesting that all the power spectra
obtained analytically in the CRWA can find their counterparts in the
exact spectrum, and the agreement is good up to $g=0.2$.

Next, let us analyze the structure of the spectrum in the CRWA in the time
scale of $\tau =2gt$. The Rabi frequency is centered around $\omega _R/2\pi =
$ $\sqrt{\alpha ^2+1}/2\pi =0.53$, independent of $g$, and therefore appear
in the same position. The Rabi oscillations in the CRWA are qualitatively
the same as those in RWA. A slight deviation from the exact ones with an
increase in the coupling strength is due to a small deviation of the CRWA
energy levels from the exact solutions.

The intrinsic oscillations from the CRTs are not captured by the RWA. The
two frequencies from intrinsic transitions within same $k$, $\omega _k^s$,
are given by Eq.~(\ref{I_close}), where two dominate frequencies come from $%
n\thickapprox \alpha ^2$ levels
\begin{equation}
\omega _k^s=\left[ 2-\frac{g^2}2+\left( -1\right) ^kg\left( \sqrt{\alpha ^2+3%
}-\sqrt{\alpha ^2+1}\right) \right] /2g.  \label{kk}
\end{equation}
Note that the third term in Eq.~(\ref{I_close}) is very small compared to
the central frequency $\omega _c=\left( 1-g^2/4\right) /g$, and is weakly
dependent on $n$. So they appear sharply in the two side of the central
frequency $\omega _c$. The other two frequencies from transitions between
different $k$'s, $\omega _k^d$, are given by Eq.~(\ref{I_apart}). Two
dominate frequencies from $n\thickapprox \alpha ^2\ $ levels are given by
\begin{equation}
\omega _k^d=\left[ 2-\frac{g^2}2+\left( -1\right) ^kg\left( \sqrt{\alpha ^2+3%
}+\sqrt{\alpha ^2+1}\right) \right] /2g.  \label{kk'}
\end{equation}
From Eq.~(\ref{I_apart}), we note however that the third term is comparable
with the central frequency $\omega _c$, and is strongly dependent on $n$.
Therefore, broad peaks appear on the two sides of the two dominate ones, $%
\omega _k^d$, in the power spectrum, as demonstrated in the exact solutions.
The peak positions from the four dominate frequencies calculated from Eqs.~(%
\ref{kk}) and (\ref{kk'}) also coincide with the exact ones.

For $g=0.06$, the spectra calculated from the CRWA are in excellent
agreement with the numerically exact ones. As the coupling strength is
increased to $g=0.15 $ and $0.2$, the aforementioned four CRWA frequencies
from Eqs.~(\ref{kk}) and (\ref{kk'}) and the Rabi peak, which are in the low
frequency regime and dominate the atomic population inversion, remain in
agreement with the exact ones. Thus, we have demonstrated that the main
features in the power spectrum of the numerically exact solutions can be
explained analytically by the CRWA. We note that a weak, broad spectral
feature, which becomes pronounced only at strong coupling, also appears at
the high frequencies, and cannot be reproduced in the CRWA.

After analyzing the structure of the wave function, it becomes clear that
the new intrinsic oscillation from the further corrections to the RWA should
take the form
\begin{eqnarray}
&\sim &g^2\alpha ^2\exp \left( -\alpha ^2\right) \sum_{k,k^{\prime
}=1,2}\sum_{n=0}^\infty \frac 1{\sqrt{n+3}}\frac{\alpha ^{2n}}{n!}\cos
\left( E_{k,n+4}t-E_{k^{\prime },n}t\right) , \\
&\sim &g^3\alpha ^2\exp \left( -\alpha ^2\right) \sum_{k,k^{\prime
}=1,2}\sum_{n=0}^\infty \frac 1{\sqrt{n+5}}\frac{\alpha ^{2n}}{n!}\cos
\left( E_{k,n+6}t-E_{k^{\prime },n}t\right) .
\end{eqnarray}
They are added to the total intrinsic oscillation gradually with the
increasing coupling. The amplitude in the $m$-th order corrections is of
order $g^m$, and therefore they only play a minor effect at weak coupling.
With the increasing coupling, they gradually gain importance. We can in
principle predict the relevant frequencies of these higher-order intrinsic
oscillations that emerge with the increasing coupling.

From Eq.~(27), it becomes immediately possible to obtain the next $4$ new
intrinsic typical frequencies in the second-order corrections to the RWA
\begin{eqnarray}
\Omega _k^s &=&\left[ 4-g^2+\left( -1\right) ^kg\left( \sqrt{\alpha ^2+5}-%
\sqrt{\alpha ^2+1}\right) \right] /2g, \\
\Omega _k^d &=&\left[ 4-g^2+\left( -1\right) ^kg\left( \sqrt{\alpha ^2+5}+%
\sqrt{\alpha ^2+1}\right) \right] /2g,
\end{eqnarray}
Note that the CRWA energy levels will still be used due to the lack
of analytical solutions in the second order corrections to the RWA.
The peak positions calculated from the equations above using the
CRWA energy are in agreement with the numerically exact ones for
$g=0.15$ and $g=0.2$, as shown in Figs.~\ref{spectrum} (b) and (c).
The small deviation from the exact spectra are attributed to the
approximate CRWA energies. Were numerically exact energy levels
used, the agreement should be much improved.

\section{summary}

In summary, using the CRWA, we have studied analytically the atomic
population inversion from an initial coherent field. The often-seen
collapse in the RWA is found to be wiped out by intrinsic
oscillations attributed to the transitions between ($k,n$) and
($k^{\prime }=k,n+2$) levels. As the intrinsic oscillations are
ubiquitous in atom-cavity coupled systems as long as the CRTs are
taken into account, it is concluded that the collapse is just an
artifact of the RWA. In addition, we have analyzed the power
spectrum of the atomic population inversion, finding the analytical
CRWA spectrum in excellent agreement with that from the exact
solutions in the ultra-strong coupling regime. As the coupling is
further increased, the main characteristic frequencies can be well
accounted for by the CRWA. Second-order corrections to the RWA are
shown to be able to explain additional features in the power
spectrum calculated from the exact solutions. Finally, our
prediction on the absence of the collapse can be checked
experimentally in the ultra-strong coupling regime.

\section{ACKNOWLEDGEMENTS}

We acknowledge useful discussion with Shi-Yao Zhu and Yu-Yu Zhang.
This work is supported by
National Basic Research Program of China (China), 2011CBA00103
, National Natural Science Foundation of China (China), 11174254, 11474256,
, and the
Singapore National Research Foundation through the Competitive Research
Programme (CRP), NRF-CRP5-2009-04.

$^{\ast }$ Corresponding author. Email:qhchen@zju.edu.cn

\appendix

\section{Derivation of the atomic population inversion}

The initial coherent state in the upper atomic level can be expanded as
\begin{equation}
|\psi (0)\rangle =\exp (-\alpha ^2/2)|\uparrow \rangle \sum_{n=0}\frac{\alpha
^n}{\sqrt{n!}}|n\rangle =|\uparrow \rangle \sum_{n=0}\beta _n|n\rangle ,
\end{equation}
In terms of CRWA eigenstates, the time dependent state in $|\uparrow
,n\rangle $ can be expressed in detail as follows. For $n=0$
\[
\sum_{k=1,2}\left[ \beta _0(c_{k0}^{(0)})^2+\beta
_2c_{k0}^{(2)}c_{k0}^{(0)}\right] e^{-iE_{k0}t},
\]
for $n=1$
\[
\sum_{k=1,2}\left[ \beta _1(c_{k1}^{(0)})^2+\beta
_3c_{k1}^{(2)}c_{k1}^{(0)}\right] e^{-iE_{k1}t}+\beta
_1(d_1)^2e^{-iE_{GS}t},
\]
and generally $n\geq 2$
\[
\sum_{k=1,2}\left[ \beta _{n-2}c_{kn-2}^{(0)}c_{kn-2}^{(2)}+\beta
_n(c_{kn-2}^{(2)})^2\right] e^{-iE_{kn-2}t}+\left[ \beta
_n(c_{kn}^{(0)})^2+\beta _{n+2}c_{kn}^{(0)}c_{kn}^{(2)}\right]
e^{-iE_{kn}t}.
\]
Then the upper part of time dependent wave function can be written as
\begin{align}
& |\psi (t)\rangle =\sum_{n=0}\sum_{k=1,2}\left[ \beta
_nc_{kn}^{(0)}c_{kn}^{(2)}+\beta _{n+2}(c_{kn}^{(2)})^2\right]
e^{-iE_{kn}t}|n+2\rangle  \nonumber \\
& +\sum_{n=0}\sum_{k=1,2}\left[ \beta _n(c_{kn}^{(0)})^2+\beta
_{n+2}c_{kn}^{(0)}c_{kn}^{(2)}\right] e^{-iE_{kn}t}|n\rangle +\beta
_1(d_1)^2e^{-iE_{GS}t}|1\rangle  \nonumber \\
& =\sum_{n=0}\sum_{k=1,2}\left[ h_{kn}e^{-iE_{kn}t}|n+2\rangle
+f_{kn}e^{-iE_{kn}t}|n\rangle \right] +\beta _1(d_1)^2e^{-iE_{GS}t}|1\rangle
,
\end{align}
where
\begin{align*}
& f_{kn}=\beta _n(c_{kn}^{(0)})^2+\beta _{n+2}c_{kn}^{(0)}c_{kn}^{(2)}=\beta
_n\left[ \frac 12+(-1)^k\frac 18\frac{(n-2\alpha ^2+2)}{\sqrt{n+1}}g-\frac
18\alpha ^2g^2\right] , \\
& h_{kn}=\beta _nc_{kn}^{(0)}c_{kn}^{(2)}+\beta _{n+2}(c_{kn}^{(2)})^2=\beta
_n\left[ -(-1)^k\frac 14\sqrt{n+2}g+\frac 18\sqrt{\frac{n+2}{n+1}}(\alpha
^2-n-1)g^2\right] .
\end{align*}
The probability to find the atom in the upper level is then obtained
\begin{align*}
P(t)& =C+\sum_{k=1,2}S_k\cos [(E_{GS}-E_{k1})t]+\sum_{n=0}\{R_n\cos
(E_{2n}-E_{1n})t+I_{1n}\cos (E_{1n}-E_{1n+2})t \\
& +I_{2n}\cos (E_{2n}-E_{2n+2})t+I_{12n}\cos (E_{1n}-E_{2n+2})t+I_{21n}\cos
(E_{2n}-E_{1n+2})t\},
\end{align*}
where
\begin{align}
& R_n=2(f_{1n}f_{2n}+h_{1n}h_{2n})=\frac 12\beta _n^2\left[ 1-(\frac{\alpha
^2}2+\frac 1{16}\frac{(n-2\alpha ^2+2)^2}{n+1}+\frac 14n+\frac 12)g^2\right]
,  \nonumber \\
& I_{1n}=2h_{1n}f_{1n+2}=\frac 12\beta _n^2\left\{ \left[ -\frac 18\frac{%
(n+4-2\alpha ^2)}{\sqrt{n+1}\sqrt{n+3}}+\frac 14\frac{(\alpha ^2-n-1)}{n+1}%
\right] \alpha ^2g^2+\frac 12\frac{\alpha ^2g}{\sqrt{n+1}}\right\} ,
\nonumber \\
& I_{2n}=2h_{2n}f_{2n+2}=\frac 12\beta _n^2\left\{ \left[ -\frac 18\frac{%
(n+4-2\alpha ^2)}{\sqrt{n+1}\sqrt{n+3}}+\frac 14\frac{(\alpha ^2-n-1)}{n+1}%
\right] \alpha ^2g^2-\frac 12\frac{\alpha ^2g}{\sqrt{n+1}}\right\} ,
\nonumber \\
I_{12n}& =2h_{1n}f_{2n+2}=\frac 12\beta _n^2\left\{ \left[ \frac 18\frac{%
(n+4-2\alpha ^2)}{\sqrt{n+1}\sqrt{n+3}}+\frac 14\frac{(\alpha ^2-n-1)}{n+1}%
\right] \alpha ^2g^2+\frac 12\frac{\alpha ^2g}{\sqrt{n+1}}\right\} ,
\nonumber \\
I_{21n}& =2h_{2n}f_{1n+2}=\frac 12\beta _n^2\left\{ \left[ \frac 18\frac{%
(n+4-2\alpha ^2)}{\sqrt{n+1}\sqrt{n+3}}+\frac 14\frac{(\alpha ^2-n-1)}{n+1}%
\right] \alpha ^2g^2-\frac 12\frac{\alpha ^2g}{\sqrt{n+1}}\right\} ,
\nonumber \\
C& =\sum_{k=1,2}\sum_{n=0}|f_{kn}|^2+|h_{kn}|^2=\frac 12+e^{-\alpha
^2}g^2\sum_{n=0}\frac{\alpha ^{2n}}{n!}\left[ -\frac 14\alpha ^2+\frac 1{32}%
\frac{(n-2\alpha ^2+2)^2}{n+1}+\frac n8+\frac 14\right] ,  \nonumber \\
S_k& =2f_{1k}d_1^2\beta _1=\frac 14e^{-\alpha ^2}\alpha ^2g^2.
\label{coeff_final}
\end{align}

The atomic population inversion is therefore easily given by
\begin{equation}
W^{CRWA}(t)=2P(t)-1,
\end{equation}
collecting all contribution up to the $g^2$ \ we have
\begin{equation}
W^{CRWA}(t)=\left( 2C-1\right) +W_{GS}(t)+W^{(\mathrm{Rabi})}(t)+W^{(I)}(t),
\label{w_tot}
\end{equation}
where
\begin{equation}
2C-1=g^2\sum_{n=0}\beta _n^2\left[ \frac 12+\frac n4-\frac{\alpha ^2}2+\frac{%
\left( n-2\alpha ^2+2\right) ^2}{16\left( n+1\right) }\right] ,
\label{const}
\end{equation}
\begin{equation}
W_{GS}=\alpha ^2g^2e^{-\alpha ^2}\cos [(2-\frac{g^2}4)t]\cos [(1-\frac{15g^2%
}{64})\sqrt{2}gt],  \label{wgs}
\end{equation}
\begin{equation}
W^{(\mathrm{Rabi})}(t)=\sum_{n=0}\beta _n^2\left[ 1-\left( \left( \frac
12+\frac n4+\frac{\alpha ^2}2+\frac{\left( n-2\alpha ^2+2\right) ^2}{%
16\left( n+1\right) }\right) g^2\right) \right] \cos (E_{2n}-E_{1n})t,
\label{wr}
\end{equation}
\begin{eqnarray}
W^{(I)}(t) &=&-\frac 12g\alpha ^2\sum_{k,k^{\prime }=1,2}\sum_{n=0}^\infty
\frac{\beta _n^2}{\sqrt{n+1}}\left( -1\right) ^{k^{\prime }}\cos \left(
E_{k,n+2}t-E_{k^{\prime },n}t\right)   \nonumber \\
&&+\alpha ^2g^2\sum_{n=0}\beta _n^2\cos \left[ \left( 2-\frac 12g^2\right)
t\right] \left\{
\begin{array}{c}
\frac{(\alpha ^2-n-1)}{n+1}\cos \left( g\sqrt{n+1}t\right) \cos \left( g%
\sqrt{n+3}t\right)  \\
-\frac 12\frac{(n+4-2\alpha ^2)}{\sqrt{n+1}\sqrt{n+3}}\sin \left( g\sqrt{n+1}%
t\right) \sin \left( g\sqrt{n+3}t\right)
\end{array}
\right\} .  \nonumber
\end{eqnarray}

\section{Approximate short-time dynamics using the saddle-point method}

By using the saddle-point method, the asymptotic expansion of integrals of
the following form is given by
\begin{equation}
F(\overline{n})=\int_0^\infty f(n)e^{\overline{n}S(n)}dn=\sqrt{-\frac{2\pi }{%
\overline{n}S^{\prime \prime }(n_0)}}\exp \left[ \overline{n}S(n_0)\right]
\left[ f(n_0)+O(\overline{n}^{-1})\right] ,  \label{saddle}
\end{equation}
where $\overline{n}$ is very large, the zeros of $S^{\prime }(n)$ are called
the saddle points of $S(n)$.

The envelope of Eq.~(\ref{I1}) can be transformed to the following integral
[c.f. Ref. (\cite{eberly1})]
\begin{eqnarray*}
F &(t)=&\sum_{n=0}^\infty \frac{\beta _n^2}{\sqrt{n+1}}\sin \left[ g\left(
\sqrt{n+1}+\sqrt{n+3}\right) t\right] , \\
&=&\int_0^\infty dn\frac{\exp \left[ -\overline{n}+n-n\ln \left( \frac n{%
\overline{n}}\right) \right] }{\sqrt{2\pi n}}\frac 1{\sqrt{n+1}}\sin \left[
g\left( \sqrt{n+1}+\sqrt{n+3}\right) t\right] ,
\end{eqnarray*}
where the Stirling's approximation was used. By  $\cos \left( x\right)
+i\sin \left( x\right) =e^{ix}$ , we have
\begin{eqnarray*}
F(t) &=&{Im}\left\{ \int_0^\infty dn\frac{\exp \left[ -\overline{n}+n-n\ln
\left( \frac n{\overline{n}}\right) +i\left( \sqrt{n+1}+\sqrt{n+3}\right)
gt\right] }{\sqrt{2\pi n}}\frac 1{\sqrt{n+1}}\right\} , \\
&=&\exp \left( -\overline{n}\right) {Im}\left( \int_0^\infty dn\frac{\exp
\left( \overline{n}S(n)\right) }{\sqrt{2\pi n\left( n+1\right) }}\right) ,
\end{eqnarray*}
where
\[
\ S(n)=\frac n{\overline{n}}-\frac n{\overline{n}}\ln \left( \frac n{%
\overline{n}}\right) +i\frac{\sqrt{n+1}+\sqrt{n+3}}{\overline{n}}gt.
\]
The zero of $S^{\prime }(n)\;$give the saddle point $n_0$
\[
\ln \left( \frac{n_0}{\overline{n}}\right) =i\left( \frac 1{\sqrt{n_0+1}%
}+\frac 1{\sqrt{n_0+3}}\right) \frac{gt}2.
\]
Considering short-time $gt/\sqrt{\bar{n}}\ll 1,$ so  $\frac{n_0}{\overline{n}%
}\rightarrow 1,\;$we have
\begin{equation}
n_0\thickapprox \bar{n}(1+\frac{igt}{\sqrt{\bar{n}}}).  \label{point}
\end{equation}
By using Eq.~(\ref{saddle}) , we have
\[
F(t)=\exp \left( -\overline{n}\right) {Im}\left( \sqrt{-\frac{2\pi }{%
\overline{n}S^{\prime \prime }(n_0)}}\exp \left[ \overline{n}S(n_0)\right]
\frac 1{\sqrt{2\pi n_0(n_0+1)}}\right) .
\]
Inserting Eq.~(\ref{point}) , we finally obtain
\begin{equation}
F(t)\thickapprox \sqrt{\frac 1{\bar{n}}}\sin (2g\sqrt{\bar{n}}t)\exp \left[ -%
\frac{\left( gt\right) ^2}2\right] .  \label{short}
\end{equation}

\end{document}